\documentclass{article}
\usepackage{subcaption,rotating,xspace}
\usepackage[linesnumbered,ruled,vlined, algo2e]{algorithm2e}
\usepackage{algorithm}

\usepackage[margin=1.0in]{geometry}

\usepackage[utf8]{inputenc}
\usepackage{amsmath}
\usepackage{amsfonts}
\usepackage{amssymb}
\usepackage{mathrsfs}
\usepackage{mathtools}
\usepackage[square,sort,comma,numbers]{natbib}
\usepackage{listings}\usepackage{color}
\usepackage{tikz}
\usepackage{bbm}
\usepackage{comment}

\definecolor{codegreen}{rgb}{0,0.6,0}
\definecolor{codegray}{rgb}{0.5,0.5,0.5}
\definecolor{codepurple}{rgb}{0.58,0,0.82}
\definecolor{backcolour}{rgb}{0.95,0.95,0.92}
 
\newcommand{\dup}{\textrm{dup}}

\newcommand{\cale}{\textrm{cal}_T}
\newcommand{\butt}{\textrm{butt}_{k}}
\newcommand{\imp}{\textrm{imp}}\renewcommand{\imp}{\Sigma}
\newcommand{\duph}{\dup^h}
\newcommand{\thep}{p}
\newcommand{\theph}{\thep^h}
\newcommand{\kk}{\kappa}

\newtheorem{thm}{Theorem}
\newtheorem{prop}[thm]{Proposition}
\newtheorem{Rmq}[thm]{Remark}
\newcommand{\be}{\begin{equation}}
\newcommand{\ee}{\end{equation}}

\newcommand{\br}{\begin{Rmq}}
\newcommand{\er}{\end{Rmq}}

\usepackage{color}

\DeclareMathOperator*{\argmax}{arg\,max}
\DeclareMathOperator*{\argmin}{arg\,min}

\newcommand{\finproof}{\rule{4pt}{6pt}}

\def\thep{F}\def\thep{p}
\def\they{\boldsymbol{y}}

\title{Beyond Surrogate Modeling: Learning the Local Volatility Via Shape Constraints\footnote{
Single-file demos  Master.html and Master.ipynb are available on [https://github.com/mChataign/Beyond-Surrogate-Modeling-Learning-the-Local-Volatility-Via-Shape-Constraints]. Note that, due to github size limitations, the file Master.html file must be downloaded locally (and then opened with a browser) to be displayed.}}
\author{Marc Chataigner\thanks{LaMME, Universit\'e d'Evry, CNRS, Universit\'e Paris-Saclay; marc.chataigner@univ-evry.fr. 
The PhD thesis of Marc Chataigner is co-funded
by the Research Initiative ``Mod\'elisation des march\'es actions, obligations
 et d\'eriv\'es'',
  financed by HSBC France under the aegis of the Europlace Institute of Finance, and by the public grant 
ANR-11-LABX-0056-LLH LabEx LMH.
}
\and Areski Cousin \thanks{Institut de Recherche en Math\'ematique Avanc\'ee, Universit\'e de Strasbourg,
7 rue Ren\'e Descartes, 67084 Strasbourg, cedex;
a.cousin@unistra.fr}\and
St\'{e}phane Cr\'{e}pey\thanks{LPSM, Universit\'e de Paris; Stephane.Crepey@lpsm.paris. The research of S. Cr\'{e}pey benefited from the support of the Chair Stress Test, RISK Management and Financial Steering, led by the French Ecole polytechnique and its Foundation and sponsored by BNP Paribas.}
\and Matthew Dixon\thanks{Department of Applied Mathematics, Illinois Institute of Technology, Chicago; matthew.dixon@iit.edu.} \and
Djibril Gueye\footnotemark[3]}

\def\sp{\,,\;}
\renewcommand\r[1]{#1}\renewcommand\r[1]{{\color{red}#1}}
\renewcommand\b[1]{{\color{blue}#1}}

\def\x{x}
\def\rqT{\int_0^T (r(s)-q(s))ds}\def\rqT{(r-q)T}
\def\eqd{:aa=}\def\eqd{=}
 
\def\reg{\mathcal{R}}
\def\IV{IV\xspace}\def\IV{implied volatility\xspace}

\begin{document}

\maketitle

\begin{abstract} We explore the abilities of
two machine learning approaches for  
 no-arbitrage interpolation of European vanilla option prices, which jointly yield the corresponding local volatility surface: a finite dimensional Gaussian process (GP) regression approach
 under no-arbitrage constraints based on prices, and a neural net (NN) approach with penalization of arbitrages based on implied volatilities.
 We demonstrate the performance of these approaches relative to the SSVI industry standard. 
 The GP approach is proven arbitrage-free, whereas arbitrages are only penalized under the SSVI and NN approaches.
The GP approach obtains the best out-of-sample calibration error and provides uncertainty quantification.
The NN approach yields a smoother local volatility and a better backtesting performance, as its training criterion incorporates a local volatility regularization term. 
%The SSVI industry standard is faster.
% The NN approach based on implied volatilities has the best
% out-of-sample and backtesting performance, but its training time is significantly longer than for the GP approach, and SSVI is the fastest. Only the GP approach provides uncertainty quantification, which is useful for the assessment of model risk.
 \end{abstract}
% \begin{abstract}
% Gaussian processes (GPs) for option pricing and greeking have emerged as novel ``kriging" methodologies for fast computations with applications in hedging and risk management. However, in general, these approaches do not enforce no-arbitrage conditions, and therefore the subsequent local volatility surface is not considered. In this article, we develop a 
% %finite dimensional 
% kriging approach, under constraints for  
%  no-arbitrage interpolation of European vanilla market prices, which jointly yields the full surface of local volatilities with uncertainty bands.  
%  We provide the experimental design parameters that are needed for competitive performance of GPs on a real dataset of SPX vanilla options. Furthermore we demonstrate the performance relative to  
%  alternative interpolation techniques including SSVI and %no-arbitrage
%  deep learning of implied volatilities with shape constraint regularization.  
%  \end{abstract}
 
 % Keywords


% \vspace{2mm}
% \noindent
% \textbf{Mathematics Subject Classification:} 
% 91B25, %Asset pricing models 
% 91G20, %Derivative securities 
% 62G08, % Nonparametric regression 
% 68Q32. %Computational learning theory

\section{Introduction}\label{sect:intro}
There have been recent surges of literature about the learning of derivative pricing functions by machine learning surrogate models, i.e.~neural nets and Gaussian processes that are respectively surveyed in \cite{ruf2019neural}  and \cite[Section 1]{crpey2019gaussian}.
% \footnote{{to which we refer the reader regarding literature review for the sake of space here;
%   see also, more recently, \citep{ludkovski2020krighedge}.}}
% investigate the application of GPs to local volatility modeling but using analytic surrogate derivatives and a focus on experimental design rather than considering the use of shape constraints in the absence of a surrogate model. .
% , for acceleration purposes (once the learning task achieved).
 There has, however, been relatively little coverage of no-arbitrage constraints when interpolating prices, and of the ensuing question of extracting the corresponding local volatility surface.
%  %with neural nets. The corresponding coverage of GPs is still scarcer. 
%  This points to methodological questions of how to effectively impose static butterfly and calendar spread no-arbitrage constraints when learning prices.
%  %and the relative merits of this approach compared to the more studied neural networks. 
%  Then there are empirical questions around relative performance and training times compared to industry best practices.
 
 Tegn{\'e}r \& Roberts \cite[see their Eq.~(10)]{tegner2019probabilistic} first attempt the use of GPs for local volatility modeling by placing a Gaussian prior directly on the  local volatility surface.  
 Such an approach leads to a
nonlinear
 least squares training loss function,
which is not obviously amenable to gradient descent (stochastic or not), so the authors {resort} to a MCMC optimization. Zheng et al. \cite{zheng2020gated}
introduce shape constraint penalization via a multi-model gated neural network, which uses an auxiliary network to fit the parameters. The gated network is interpretable and lightweight, but the training is expensive and there is no guarantee of no-arbitrage. They do not consider the local volatility and the associated regularization terms, nor do they assess the extent to which no-arbitrage is violated in a test set.

Maatouk \& Bay \cite{maatouk2017gaussian}
introduce finite dimensional approximation of Gaussian processes (GP) for which shape constraints are straightforward to impose and verify.  
 Cousin et al. \cite{ARES} 
 apply this technique   
 to ensure arbitrage-free and error-controlled yield-curve and CDS curve interpolation.

In this paper, we propose
an arbitrage-free GP option price interpolation, which
jointly yields the corresponding local volatility surface, with
 uncertainty quantification. Another contribution of the paper is to introduce a neural network approximation of the \IV  surface, penalizing arbitrages on the basis of
the Dupire formula, which is also used for extracting 
the corresponding local volatility surface. This is all evidenced on an SPX option dataset.

Throughout the paper we
 consider European puts on a stock (or index) $S$ with dividend yield $q$, in an economy with interest rate term $r$, with $q$ and $r$ constant in the mathematical description and deterministic in the numerics.

Given any rectangular domain of interest in time and space,
we tacitly rescale the inputs so that the domain becomes $\Omega\eqd [0,1]^2$. This rescaling avoids any one independent variable dominating over another during any fitting of the market prices.

\section{Gaussian process regression for learning arbitrage-free price surfaces} \label{ref:GP-prices}
{\def\co{\boldsymbol{\kappa}}\def\co{c}
\def\veco{\mathbf{K}}\def\veco{\boldsymbol{\co}}
\def\maco{\mathbb{K}}\def\maco{\mathbf{C}}
\def\cC{\mathacal{M}}\def\cC{\mathcal{C}}\def\cC{\mathcal{I}}

We denote by $P_* (T,K)$ the time-0 market price of the
put with maturity $T$ and strike $K$ on $S$,
observed for a finite number of pairs $(T,K)$. 
Our first goal is to construct, by Gaussian process regression,
%or neural net interpolation, 
an arbitrage-free
and continuous put price surface $P:\mathbb{R}_+\times \mathbb{R}_+\rightarrow\mathbb{R}_+$, 
interpolating $P_*$ up to some error term,
and to retrieve the corresponding local volatility surface $\sigma(\cdot ,\cdot)$ by
the   
Dupire formula.  
 
In terms of the reduced prices $\thep(T,k) \eqd 
e^{q T}
%e^{\int_{0}^{T} q(t) dt}
P(T,K),$
where $k= K 
e^{-  (r-q) T}
%e^{-\int_{0}^{T} (r(t) - q(t)) dt}
,$
the Dupire formula  \cite{Dupire1994} reads (assuming $p$ of class $\mathcal{C}^{1,2}$ on $\{T>0\}$):
%(see e.g.~the appendix in \cite{Chataigner2020Deep}) 
\begin{equation}\label{eq:dupk}
\frac{\sigma^2(T,K)}{2} = \frac{ \partial_T \thep(T,k)}{k^2 \partial_{k^2}^2 \thep(T,k)} =: \dup(T,k).
\end{equation}   
Obviously, for this formula to be meaningful, its output must be nonnegative, which holds if the interpolating map $\thep$
exhibits nonnegative derivatives w.r.t.~T and second derivative w.r.t.~k, i.e.
\begin{equation}\label{e:na}
\partial_T \thep(T,k)\ge 0\sp  \partial_{k^2}^2 \thep(T,k)\ge 0,
\end{equation}

In this section,
% we construct reduced put price surfaces $(T,k)\mapsto \thep(T,k)$ satisfying the conditions \eqref{e:na} from $n$ noisy observations $\they=[y_{1},\dots,y_{n}]^\top$
% of function $\thep$ at input points  $\mathbf{\x}=[x_{1},\dots,x_{n}]^{\top}$.  The input points $x_{i}=(T_{i}, k_{i})$ correspond to observed maturities and strikes. The  market fit condition is written as
% \be\label{e:GM} \they = \thep(\mathbf{\x}) + \boldsymbol{\varepsilon},  \ee where $\thep(\mathbf{\x})\eqd [\thep(x_{1}),\dots,\thep(x_{n})]^{\top}$ is the vector composed of {the bid and ask put prices} at the observation points. The  additive noise term $\boldsymbol{\varepsilon}=[\varepsilon_{1}, \dots,\varepsilon_{n}]^{\top}$ is assumed to be a zero-mean Gaussian vector, independent from $\thep(\mathbf{\x})$, and with an homoscedastic covariance matrix given as $\varsigma^{2} I_{n}$, where $I_n$ is the identity matrix of dimension $n$. \b{Note that bid and ask prices are considered here as (noisy) replications at the same input location.}
% \r{
% In what follows, 
we consider a zero-mean Gaussian process prior on the mapping $\thep=\thep(x)_{x\in \Omega}$ with correlation function $\co $ given, for any $x=(T,k), x'=(T',k') \in \Omega$, by
\be\label{e:ke}  \co (x,x')=\sigma^2 \gamma_{T}(T-T', \theta_T)\gamma_{k}(k-k', \theta_k).\ee
 Here $(\theta_T, \theta_{k} )=\theta$ 
  %=:\boldsymbol{\theta}
  and $\sigma^2$ correspond to  length scale and variance hyper-parameters of the kernel function $\co $, whereas the functions $\gamma_{T}$ and $\gamma _{k}$ are kernel correlation functions. 
 
Without consideration of the conditions \eqref{e:na}, (unconstrained)  prediction and uncertainty quantification are made using the conditional distribution $\thep\; \vert \; \thep(\mathbf{\x}) + \boldsymbol{\varepsilon}=\they $, where
$\they=[y_{1},\dots,y_{n}]^\top$ are
$n$ noisy observations 
of the function $\thep$ at input points  $\mathbf{\x}=[x_{1},\dots,x_{n}]^{\top}$,  corresponding to observed maturities and strikes  $x_{i}=(T_{i}, k_{i})$;
% The  market fit condition is written as
% \be\label{e:GM} \they = \thep(\mathbf{\x}) + \boldsymbol{\varepsilon},  \ee where $\thep(\mathbf{\x})\eqd [\thep(x_{1}),\dots,\thep(x_{n})]^{\top}$ is the vector composed of {the bid and ask put prices} at the observation points. 
the  additive noise term $\boldsymbol{\varepsilon}=[\varepsilon_{1}, \dots,\varepsilon_{n}]^{\top}$ is assumed to be a zero-mean Gaussian vector, independent from $\thep(\mathbf{\x})$, and with an homoscedastic covariance matrix given as $\varsigma^{2} I_{n}$, where $I_n$ is the identity matrix of dimension $n$. Note that bid and ask prices are considered here as (noisy) replications at the same input location.

 \subsection{Imposing the no-arbitrage conditions}\label{sub:na}
To deal with the
%inequality 
constraints \eqref{e:na},
 we adopt the  solution of Cousin et al. \cite{ARES} that consists in constructing a finite dimensional approximation $\theph$ of the Gaussian prior $\thep$ for which these constraints can be imposed in the entire domain $\Omega$ with a finite number of checks.
 One then recovers the (non Gaussian) constrained posterior distribution by sampling a truncated Gaussian process.
 \begin{Rmq}
%  In the original infinite dimensional situation, testing the inequality constraints on the entire input domain would require an infinite number of checks.   
% With the computation of the local volatility surface in mind,
Switching to a finite dimensional approximation can also be viewed as a form of regularization, which is also required to deal with the ill-posedness of the
(numerical differentiation) Dupire formula.
 \end{Rmq}
 
We first consider a discretized version of the (rescaled) input space $\Omega=[0,1]^2$ 
 as a regular grid 
 $%\Omega^h\eqd 
 ({\imath}h)_{\imath}$, where $\imath=(i,j)$,
 %$\Omega^h\eqd (\chi_{\imath})_{\imath}$, where $\imath=(i,j)$ and $ \imath h= \imath h$, 
for a suitable mesh size $h$ and  indices $i,j$ ranging from 0 to $1/h$ (taken in $\mathbb{N}^\star$).
For each knot $ \imath =(i,j)$, we introduce the hat basis functions $\phi_\imath$ 
with support $[(i-1)h,(i+1)h]\times[(j-1)h,(j+1)h]$
given, for $x=(T,k)$, by
$$\phi_{\imath}(x) =\max(1-\frac{| T- ih |}{h},0)\max(1-\frac{| k-j h|}{h},0) .$$
 
We take $V=H^1(\Omega)\eqd \{u\in L_2(\Omega) : D^{\alpha}u\in L_2(\Omega),|\alpha|\leq1\}$, where $D^{\alpha}u$ is a weak derivative of order $|\alpha|$,
as the space of (the realizations of) $\thep$. Let $V^h\subset V$ denote the finite dimensional linear subspace spanned by the $M$ linearly independent basis functions $\phi_{\imath}$.
The (random) surface $\thep$ in $V$ is projected onto $V^h$ as
\begin{equation} \label{FN}
\theph(x)=\sum_{\imath} \thep( \imath h)\phi_{\imath}(x), ~\forall x \in \Omega.
\end{equation}  
% $\thep^h\in V^h$ is a bilinear quadrilateral finite element approximation of the values of $\thep$ at knots $( \imath h)_{\imath}$.
If we denote $\varrho_{\imath}=\thep( \imath h)$,  
then $\boldsymbol{\varrho}= (\varrho_{\imath})_{\imath}$
is a zero-mean Gaussian column vector (indexed by $\imath$) with $M \times M$ covariance matrix $\Gamma^h$ such that
$\Gamma^{h}_{\imath,\jmath}= \co ( \imath h,\jmath h)$,  
for any two grid nodes $\imath$ and $\jmath$. 
Let $\boldsymbol{\phi}(x)$ denote the vector of size $M$ given by $\boldsymbol{\phi}(x)= (\phi_{\imath}(x))_{\imath}.$
The equality \eqref{FN}  can be rewritten as $\theph(x)= \boldsymbol{\phi}(x)\cdot\boldsymbol{\varrho}.$ Denoting by  
$\theph(\mathbf{\x})\eqd [\theph(x_{1}), \dots, \theph(x_{n})]^{\top}$
and by
$%\mathbf{\Phi }\eqd 
\mathbf{\Phi }(\mathbf{\x})$ the $n\times M $  matrix of basis functions where each row $\ell$ corresponds to the vector $\boldsymbol{\phi}(x_{\ell})$, one has  $\theph(\mathbf{\x})= \mathbf{\Phi }(\mathbf{\x})\cdot \boldsymbol{\varrho}  .$ 
By application of the results of \cite{maatouk2017gaussian}:
\begin{prop}    \label{PropPN}	{\rm (i)}
	The finite dimensional process $\theph$ converges  uniformly to $\thep$ on 	$\Omega$ as $h \rightarrow 0$, 
	%$h_k\rightarrow 0$ and $h_T\rightarrow 0$,
	almost surely,
%	\item[{\bf(ii)}]

\noindent
{\rm 	(ii)}
$\theph(T,k)$ is a nondecreasing  function  of $T$ if and only  if $\varrho_{i+1, j} 	\geq \varrho_{i, j},\forall (i,j)$, 
	%\item[{\bf(iii)}] 

	\noindent
{\rm 	(iii)}
	$\theph(T,k)$ is a convex function of $k$ if and only if  $\varrho_{i, j+2}-\varrho_{i, j	+1} \geq \varrho_{i, j+1}-\varrho_{i, j},\forall (i,j) $.~\finproof
\end{prop}

\noindent
In view of (i),   denoting by $\cC$ 
% as the convex set of inequality constraints, i.e., $\cC$ is 
the set  of 2d continuous positive functions which are nondecreasing in $T$ and convex in $k$, 
we choose as constrained GP metamodel for the put price surface
% our construction problem (we denote it ($\mathcal{P}$)) consists in finding 
the law of  $\theph$ conditional on
\begin{eqnarray*} \label{}  \left\{\begin{array}{ll}
\theph(\mathbf{\x})+ \boldsymbol{\varepsilon} =\they   \\
\theph \in \cC.
\end{array}\right.
\end{eqnarray*}  
In view of (ii)-(iii), 
$  \theph \in \cC\; \text{ if and only if} \;  \boldsymbol{\varrho}\in  \cC^h  ,$ where $\cC^h $ corresponds to the set of
($\imath$ indexed)
vectors 
$\boldsymbol{\rho}= (\rho_{\imath})_{\imath}$
such that
$\rho_{i+1, j} 	\geq \rho_{i, j}$ and 
$\rho_{i, j+2}-\rho_{i, j	+1} \geq \rho_{i, j+1}-\rho_{i, j}$ $\forall (i,j) $.
Hence, our GP metamodel
for the put price surface 
can be reformulated as
the law of $\boldsymbol{\varrho}$ conditional on
\begin{eqnarray}\label{e:conditi}\left\{\begin{array}{ll}
  \boldsymbol{\Phi} (\mathbf{\x})  \cdot \boldsymbol{\varrho} + \boldsymbol{\varepsilon}=\they\\  
\boldsymbol{\varrho}  \in \cC^h . 
 \end{array}\right.
\end{eqnarray}  
 
 \subsection{Hyper-parameter learning}
Hyper-parameters consist in the length scales $\theta$ and the variance parameter $\sigma^2$ in \eqref{e:ke}, as well as the noise variance $\varsigma$. 
Up to a constant, the 
so called marginal log likelihood of $\boldsymbol{\varrho}$ at $\lambda= [\theta, \sigma, \varsigma]^\top$ can be expressed as (see e.g. \cite[Section 15.2.4, p. 523]{Murphy12}):
\begin{eqnarray*}
%\nonumber    
\mathcal{L}(\lambda) =-\frac{1}{2} \boldsymbol{\they}^{\top} \big(\boldsymbol{\Phi} (\mathbf{\x})\Gamma^{h}  \boldsymbol{\Phi} (\mathbf{\x})^{\top} + \varsigma^{2} I_{n}\big)^{-1} \boldsymbol{\they } -\frac{1}{2} \log\Big(\det\big( \boldsymbol{\Phi} (\mathbf{\x})\Gamma^{h} \boldsymbol{\Phi} (\mathbf{\x})^{\top} + \varsigma^{2} I_{n} \big)\Big). %\\\label{ML} -\frac{n}{2} \log(2\pi). 
\end{eqnarray*}   
  We  maximize $\mathcal{L}$ for learning the hyper-parameters $\lambda$  (MLE estimation).  
  
  \begin{Rmq}
The above expression does not take into account the inequality constraints in the estimation. However,  Bachoc et al.~\cite[see e.g.~their Eq.~(2)]{bachoc2019maximum}
argue (and we observed
%including
empirically) that, unless the sample size is very small,
% , for large sample sizes{??}, 
conditioning by the constraints 
%only 
significantly increases the computational burden with negligible impact on the MLE.
  \end{Rmq}

\subsection{The most probable response surface and measurement noises\label{ss:measnoise}}
% The 
% maximum a posteriori probability
% estimate (MAP) 
% $\hat{p}^h$
% of $\theph$ given the constraints satisfies the constraints on the entire domain of interest and corresponds to the most likely surface. Its expression is given in \cite{ARES}: 
% \begin{equation}\label{e:maprho}
% \hat{p}^h (x)\eqd  \sum_{\imath}  \hat{\boldsymbol{\varrho}}_{\imath}\phi_{\imath}(x)  ,
% \end{equation} 
% where $\hat{\boldsymbol{\varrho}}  
% $ is the MAP of  
%  $\boldsymbol{\varrho}$.
% In order to also identify
%  the locations $x$ of the most likely arbitrages in the data and to quantify the latter,
%  as the locations of the largest noises and their values,
We compute the joint MAP $ (\hat{\boldsymbol{\rho}},  \hat{\boldsymbol{e}})$ of the truncated Gaussian vector $\boldsymbol{\varrho}$ and of the Gaussian noise vector $\boldsymbol{\varepsilon}$,
$$ {(\hat{\boldsymbol{\rho}},  \hat{\boldsymbol{e}})}=
\underset{(\boldsymbol{\rho},  \boldsymbol{e})}{\argmax} \;\rm{Prob} \left(  \boldsymbol{\varrho} \in [\boldsymbol{\rho}, \boldsymbol{\rho}+d\boldsymbol{\rho}],   \boldsymbol{\varepsilon}\in [\boldsymbol{e}, \boldsymbol{e} +d\boldsymbol{e}] \mid  \boldsymbol{\Phi} (\mathbf{\x})  \cdot \boldsymbol{\varrho} + \boldsymbol{\varepsilon}  = \boldsymbol{y},\, \boldsymbol{\varrho}  \in \cC^h  \right)
$$
(for the probability measure Prob underlying the GP model).
As $({\boldsymbol{\varrho}}, {\boldsymbol{\varepsilon}})$ is Gaussian centered with block-diagonal covariance matrix 
with blocks $\Gamma^h$ and $\varsigma^{2} I_{n},$
this implies that the MAP
%the mode 
{($\hat{\boldsymbol{\rho}},  \hat{\boldsymbol{e}}$)}  is a solution to the following quadratic  problem :
\begin{equation}  \label{MAPSolver}
\underset{ \boldsymbol{\Phi} (\mathbf{\x}) \cdot \boldsymbol{\rho} + \boldsymbol{e} = \boldsymbol{y},\, \boldsymbol{\rho}  \in \cC^h  }{\argmin} \;  \left(  \boldsymbol{\rho}^\top (\Gamma^h)^{-1} \boldsymbol{\rho} + \boldsymbol{e}^\top (\varsigma^{2} I_{n})^{-1} \boldsymbol{e} \right).
\end{equation}
We define the most probable measurement noise to be {$ \hat{\boldsymbol{e}}$} and the most probable response surface  $\hat{p}^h(\mathbf{\x})\eqd   \boldsymbol{\Phi} (\mathbf{\x}) \cdot {\hat{\boldsymbol{\rho}}}$.
Distance to the data can be an effect of arbitrage opportunities within the data and/or misspecification / lack of expressiveness of the kernel.

%(see \cite{\Thetailliams2006gaussian})
 \subsection{Sampling finite dimensional Gaussian processes under shape constraints}
%  In view of
% \eqref{e:conditi}, our construction of the put price surface consists in sampling $\boldsymbol{\varrho}$
% given $\boldsymbol{\Phi} (\mathbf{\x})  \cdot \boldsymbol{\varrho} + \boldsymbol{\varepsilon}=\they$, truncated on $\cC^h$.   
The conditional distribution of $\boldsymbol{\varrho} \mid \boldsymbol{\Phi} (\mathbf{\x})\cdot\boldsymbol{\varrho}+\boldsymbol{ \varepsilon}=\they$ is multivariate Gaussian with mean $\boldsymbol\eta_{\mathbf{y}}(\mathbf{\x})$ and covariance matrix 
$\maco_{\mathbf{y}}(\mathbf{\x})$ such that  
\begin{eqnarray}\label{Mean}
&&\boldsymbol\eta_{\mathbf{y}}(\mathbf{\x})=\Gamma^h \boldsymbol{\Phi} (\mathbf{\x})^\top(\boldsymbol{\Phi} (\mathbf{\x})\Gamma^h\boldsymbol{\Phi} (\mathbf{\x})^\top +\varsigma^{2} I_{n})^{-1}\they
\\\label{Cov}
&&\maco_{\mathbf{y}}(\mathbf{\x})=\Gamma^h \boldsymbol{\Phi} (\mathbf{\x})^\top(\boldsymbol{\Phi} (\mathbf{\x})\Gamma^h\boldsymbol{\Phi} (\mathbf{\x})^\top +\varsigma^{2} I_{n})^{-1} \boldsymbol{\Phi} (\mathbf{\x})\Gamma^h . 
\end{eqnarray}
 In view of
 \eqref{e:conditi},
we  thus face the problem of sampling from this truncated multivariate Gaussian distribution, which we do by Hamiltonian Monte
Carlo,
%(see \cite{lopez2018finite}), 
using the 
MAP
$\hat{\boldsymbol{\varrho}}$
of  $\boldsymbol{\varrho}$ as the initial vector (which must verify the constraints) in the algorithm.

\subsection{Local volatility}

Due to the shape constraints and to the ensuing finite-dimensional approximation with  basis functions of class $\mathcal{C}^{0}$ (for the sake of Proposition \ref{PropPN}), $\theph$ is not differentiable.
Hence, exploiting GP 
derivatives analytics, 
as done  for the mean in \cite[cf.~Eq.~(10)]{crpey2019gaussian} and also for the covariance in \cite{ludkovski2020krighedge}, 
is not possible for deriving the corresponding local volatility surface here. %\b{To address this issue, we formulate a weak form of the Dupire equation and construct the local volatility surface approximation $\duph \in V^h$ using a bilinear quadrilateral finite element (FE) method, where $V^h$ is a finite dimensional linear subspace spanned by bilinear basis functions over quadrilateral elements. This gives rise to a linear system with a nine-stencil stiffness  matrix and a 6-stencil RHS matrix which is solved with a preconditioned GMRES solver.} 
Computation of derivatives involved in the Dupire formula is implemented by finite differences with respect to a coarser grid (than the grid of basis functions).
Another related solution would be to formulate a weak form of the Dupire equation and construct a local volatility surface approximation using a 
%bilinear quadrilateral 
finite element method. % from the weak form of the Dupire equation.
% https://github.com/
% \noindent
% mChataign/DupireNN 
%for the detail.

See Algorithm 1 %\ref{algo:GP}
for the main steps of the GP approach.
 
\begin{comment}`
 %(see appendix for further details)
$A$ can be viewed as a combination of time weighted local differencing over $\boldsymbol{\varrho}$ w.r.t. $\kk$ and subsequent averaging of $\duph$ over the four elements in each quadrant. We further note that the linear system in Eq.~\ref{eq:modstiffness} provides the opportunity for further regularization techniques such as, for examples, preconditioners (see e.g.~\cite{demmel97}).
\end{comment}
 
\begin{comment}
In order to constrain the derivatives to be greater than zero during fitting, it is necessary to evaluate the weak form of the second derivative of $\theph$ w.r.t. $k$. The weak form of the first derivative of $\theph$ w.r.t. $T$ is equivalent to the pointwise derivative because $\theph\in V^h\subset H^1(\Omega)$. Returning to grid indices $(i,j)$ we see that

$$\partial_T \theph(T,k)=\sum_{ij} \phi'_i(T)\phi_j(k)\varrho_{ij}=\frac{1}{h_k}(\varrho_{i+1,j}-\varrho_{i,j}), (T,k)\in[T_i,T_{i+1}]\times[k_j,k_{j+1}]$$
is just the 2-stencil corresponding to a forward difference over any element. Hence we may freely choose where to evaluate finite differences in each element.
The weak second derivative

$$-\int_{k=k_{min}}^{k_{max}}\theph(T,k)v^h(T,k)dk=-\sum_{\i}\varrho_{\i}\int_{k_{j-1}}^{k_{j+1}}\partial_k\phi_{\i}(T,k)\partial_k\phi_{\j}(T,k)dk=\sum_{\i}K_{\j,\i}\varrho_{\i},$$
can be written as $K\mathbf{\varrho}$, where the stiffness matrix is a 3-stencil corresponding to a second order finite difference operator.
\end{comment}
 
\begin{algorithm}[H]
\SetAlgoLined
\KwData{Put price training set $\thep_\star$}
\KwResult{$M$ realizations of the local volatility surface $\{\text{dup}^h_i\}_{i=1}^M$}
$\hat{\lambda}\leftarrow$ Maximize the marginal log-likelihood of the put price surface $p^h$ w.r.t. $\lambda$ \tcp*[h]{Hyperparameter fitting}\;

$(\hat{\boldsymbol{\rho}},\hat{\boldsymbol{e}}) \leftarrow$  Minimize quadratic problem \ref{MAPSolver} based on $\hat{\lambda}$ \tcp*[h]{Joint MAP estimate}\;

$\hat{\boldsymbol{\rho}}\rightarrow $ Initialize a Hamiltonian MC sampler\;

$\thep^h_1, \dots, \thep^h_M \leftarrow$ Hamiltonian MC Sampler  \tcp*[h]{Sampling price surfaces} \;

$\text{dup}^h_i\leftarrow$ Finite difference approximation using each $\thep^h_i,~i:=1\rightarrow M$\; %\tcp*[h]{Local volatility realizations}

\caption{The GP algorithm for local volatility surface approximation.}
\end{algorithm}

\section{Neural networks implied volatility  metamodeling}\label{sect:nets}

Our second goal is to use neural nets (NN) to construct an \IV  (IV) put surface
$\imp :\mathbb{R}_+\times \mathbb{R}\rightarrow\mathbb{R}_+$, interpolating \IV  market quotes $\imp _*$ up to some error term, both being stated in terms of a put option maturity $T$ and log-(forward) moneyness $\kappa=\log(\frac{k}{S_0})=\log{\left( \frac{K}{S_0} \right)}-\rqT
$.
The advantage of using implied volatilities rather than prices (as previously done in \cite{Chataigner2020Deep}), both being in bijection via the Black-Scholes put pricing formula as well known, is
their lower variability, hence better performance as we will see.
% that it provides a scale-free market convention for pricing an option.
% $$P_{BS}(S, t, K, r, \imp (S,t)) = P(S, t, K, r)$$

The corresponding local volatility surface $\sigma$ is given by the following local volatility implied variance formula, i.e.~the Dupire formula stated in terms of the implied total variance\footnote{This 
%formula
follows from 
the  
Dupire formula 
by simple transforms detailed in \cite[p.13]{gatheral2011volatility}.} $\Theta(T,\kappa) \eqd  \imp ^2(T,\kappa) T$
(assuming $\Theta$ of class $\mathcal{C}^{1,2}$ on $\{T>0\}$): 
\begin{small}
\begin{equation}\label{gatheral}
\sigma^2(T,K) = \frac{\partial_{T}\Theta}{1 - \frac{\kappa}{\Theta} \partial_{\kappa}\Theta + \frac{1}{4}\left( -\frac{1}{4} - \frac{1}{\Theta} + \frac{\kappa^2}{\Theta^2} \right) (\partial_{\kappa}  \Theta)^2 + \frac{1}{2}\partial_{\kappa^2}\Theta} (T,\kappa)=:
\frac{\cale(\Theta)}{\butt(\Theta)} 
(T,\kappa).
\end{equation}  
\end{small}

%(with $\lambda=0$ in the nonpenalized cases)
We use a feedforward NN with weights $\mathbf{W}$, biases $\mathbf{b}$ and smooth activation functions for parameterizing the \IV and total variance, which we denote by
$$\Sigma=\Sigma_{\mathbf{W},\mathbf{b}}\sp\Theta=\Theta_{\mathbf{W},\mathbf{b}}. $$
The terms $\cale(\Theta_{\mathbf{W},\mathbf{b}})$ and $\butt(\Theta_{\mathbf{W},\mathbf{b}})$ are available analytically, by automatic differentiation,
which we exploit below
to penalize calendar spread arbitrages, i.e.~negativity of $\cale(\Theta )$, and butterfly arbitrage, i.e.~negativity of $\butt(\Theta )$.
% See \cite[Theorem 2.9]{roper2010arbitrage} for a sufficient set of static no-arbitrage conditions on $\sqrt{\Theta}$.

The training of NNs is a non-convex optimization problem and hence does not guarantee convergence to a global optimum. 
We must therefore guide the NN optimizer towards a local optima that has desirable properties in terms of interpolation error and arbitrage constraints.
This motivates the introduction of an arbitrage penalty function into the loss function to select the most appropriate local minima. An additional challenge is that  maturity-log moneyness pairs
with quoted 
option prices  
are unevenly distributed and the NN may favor fitting to a cluster of quotes to the detriment of fitting isolated points. To remedy this non-uniform data fitting problem, we re-weight the observations by the Euclidean distance between neighboring points.
More precisely, given $n$ observations $\chi_i=(T_i,\kappa_i)$ of   maturity-log moneyness pairs and of the corresponding
market implied 
%total variance $\Theta _* (\chi_i)$,
volatilities $\Sigma _* (\chi_i)$, we construct the $n\times n$ distance matrix with general term
$d(\chi_i,\chi_j) = \sqrt{\left(T_j - T_i\right)^2 + \left(\kappa_j - \kappa_i\right)^2}.$
We then define the loss weighting $w_i$ for each point $\chi_i$ as the distance $w_i = \min\limits_{j, j \neq i} d(\chi_i,\chi_j).$ with the closest point.
These modifications aim at reducing error for any isolated points. 
In addition, in order to avoid linear saturation of the neural network, we apply a further log-maturity change 
of variables (adapting the partial derivatives accordingly).

Learning the weights $\mathbf{W}$ and biases $\mathbf{b}$ to the data subject to no arbitrage soft constraints (i.e.~with penalization of arbitrages)
%on any given day,
%(for which calibration is a prerequisite),
then takes the form of the following (nonconvex) loss minimization problem:
\be \label{e:loss1}
%\displaystyle 
%(\hat{\mathbf{W}},\hat{\mathbf{b}})\in
\argmin_{\mathbf{W},\mathbf{b}}\;\;
% \thep(\mathbf{W},\mathbf{b}), \\
% %\mathcal{L}(y,\haty )& \eqd & ||y-\haty ||_2^2,\\
% \thep(\mathbf{W},\mathbf{b}) &= &
\sqrt{\frac{1}{n} 
\sum_i  
%\mathcal{L}
\left(  {w_i   \frac{\imp_{\mathbf{W},\mathbf{b}}(\chi_i ) -   {\imp }_* (\chi_i )}{{\imp }_* (\chi_i )} } \right)^2} +   \frac{\mu_w}{m}\sum_{\xi\in\Omega_h}\lambda^{\sf T} \reg(\Theta_{\mathbf{W},\mathbf{b}})(\xi  ),
\ee
where $\lambda=[\lambda_1,\lambda_2,\lambda_3]^{\top}\in \mathbb{R}_+^3 $
% ${\imp  }\eqd  \imp _{\mathbf{W},\mathbf{b}}$ and $g\eqd 
% g_{\mathbf{W},\mathbf{b}}$ is a regularization penalty vector :
and
\begin{eqnarray*}
\reg(\Theta )\eqd [
 \cale^- ( \Theta ) ,    \butt^-(\Theta )  ,
 \big( \frac{\cale}{\butt}(\Theta )  -\overline{a} \big )^+
  +  \big( \frac{\cale}{\butt}(\Theta ) -\underline{a} \big)^-
]^{\top}
\end{eqnarray*}
is a regularization penalty vector evaluated over a penalty grid ${\Omega}_h$ with $m$ nodes as detailed below.
 The error criterion is calculated as a root mean square error on relative difference, so that it does not discriminate high or low implied volatilities.
 The first two elements in the penalty vector $\reg(\Theta )$ favor the no-arbitrage conditions \eqref{e:na} and the third element favors desired lower and upper bounds 
$0<\underline{a}<\overline{a}$ (constants or functions of $T$)
  on the estimated local variance $\sigma^2(T,K)$. 
  In order to adjust the weight of penalization, we multiply our penalties by the weighting mean $\mu_w :=\frac{1}{m} \sum\limits_{i} w_i$.
Suitable values of
the ``Lagrange multipliers"
$\lambda,$ ensuring the right balance between fit to the market implied volatilities and the constraints, is then obtained by grid search. 
   Of course a soft constraint
 (penalization) approach does not fully prevent arbitrages. However, for large $\lambda$,  arbitrages are extremely unlikely to occur, except perhaps very far from $\Omega$. With this in mind,  
 we use a penalty grid ${\Omega}_h$ 
that
 extends well beyond the 
 %unit square 
 domain of the IV interpolation.  This is intended so that the penalty term penalizes arbitrages outside of the domain used for IV Interpolation.

 See Algorithm 2 for the pseudo-code of the NN approach.
\begin{algorithm}[H]
\setcounter{AlgoLine}{0}
\SetAlgoLined
\KwData{Market implied
volatility surface
$\Sigma _* $} 
\KwResult{The local volatility surface $\sqrt{\frac{\cale}{\butt}(\Theta_{\hat{\mathbf{W}},\hat{\mathbf{b}}}) }$ }
$(\hat{\mathbf{W}},\hat{\mathbf{b}})\leftarrow$ Minimize the
penalized training loss \eqref{e:loss1} w.r.t. $(\mathbf{W},\mathbf{b})$;

$\sqrt{\frac{\cale}{\butt}(\Theta_{\hat{\mathbf{W}},\hat{\mathbf{b}}})} \leftarrow$ AAD differentiation of the trained NN implied vol. surface\;
 %\tcp*[h]{Predict the local volatility surface}\;
\caption{The NN-IV algorithm for local volatility surface approximation.}
\end{algorithm}

\section{Numerical results}\label{sect:numerical}
 \subsection{Experimental design}
 %\subsection{Data}
Our training set is prepared using SPX European puts with different available strikes and maturities ranging from 0.005 to 2.5 years, listed on 18th May 2019, with $S_0=\$2859.53$. Each contract is listed with a bid/ask price and an implied volatility corresponding to the mid-price.  The associated interest rate is constructed from US treasury yield curve and dividend yield curve rates are then obtained from call/put parity applied to the option market prices and forward prices.
We preprocess the data by removing the shortest maturity options, with $T<0.055$, and the numerically inconsistent observations 
for which the gap between the listed implied volatility and the implied volatility calibrated from mid-price with our interest/dividend curves
exceeds
% whose mid-prices have \r{a recalibration error}\footnote{Recalibration error is the gap between the listed implied volatilities and the implied volatility calibrated from mid-price with our interest/dividend curves.} greater than 
5\% of the listed implied volatility.
% The latter avoids including prices in the dataset which are not correctly inverted by a bissection algorithm solver for the Black-Scholes equation such as deep out-of-the-money contracts. 
But we do not remove arbitrable observations.
The preprocessed training set is composed of
 1720 market put prices.   
 The testing set consists of a disjoint set of 1725 put prices.
 %(with different strike-maturity pairs than in the training set). 

All results for the GP method are based on using Matern $\nu=5/2$ kernels over a $[0,1]^2$ domain with fitted kernel standard-deviation hyper-parameter $\hat{\sigma}=185.7611$, length-scale hyper-parameters $\hat{\theta}_k=0.3282$ and $\hat{\theta}_T=0.2211$, and homoscedastic noise standard deviation, $\hat{\varsigma}=0.6876$.\footnote{When re-scaled back to the original input domain, the fitted length scale parameters of the 2D Matern $\nu=5/2$ are  $\hat{\theta}_k=973.1901$ and $\hat{\theta}_T=0.5594$.}
 The grid of basis functions for constructing the finite-dimensional process $\theph$ has $100$ nodes in the modified strike direction and $25$ nodes in the maturity direction. 
The Matlab interior point convex algorithm
\verb|quadprog|
 is used to solve the MAP quadratic program
 \ref{MAPSolver}.

Regarding the NN approach, we use a three layer architecture similar to the one based on prices
(instead of implied volatilities in Section \ref{sect:nets})
in \cite{Chataigner2020Deep}, to which we refer the reader for implementation details.
We use a penalty grid ${\Omega}_h$ 
with $m=50\times 100$ nodes.
 In the moneyness and maturity coordinates, the domain of  
 the penalty grid is
$[0.005,10]\times[0.5,2] $.

\subsection{Arbitrage-free SVI}

We benchmark the machine learning results with the
industry standard provided by the arbitrage free stochastic volatility inspired (SVI) model of \cite{gatheral2014arbitrage}. 
% Given SVI parameters $\theta_{\text{SVI}} = \left\{a,b,\varrho,m,\sigma \right\}$ the implied total variance is defined as the following :
% \begin{equation}\label{eq:svi}
%   \Theta(k, \theta_{\text{SVI}}) = a + b \left\{ \varrho (k - m) + \sqrt{\left( k - m \right)^2 + \sigma^2} \right\}.
% \end{equation}
Under the ``natural 
%SVI
parameterization" 
$%\theta_
{\text{SVI}} = 
(\Delta, \mu, \rho, \omega, \zeta )$, the implied total variance is given, for any fixed $T$, by 
%(in our former modified strike variable $k$)
\begin{equation}\label{eq:naturalSvi}
    \Theta_{\text{SVI}}(\kappa)
    % \Theta(\kappa,\theta_{\text{SVI}}) 
    = \Delta + \frac{\omega}{2} \left( 1 + \rho (\kappa - \mu) \zeta + \sqrt{  (\zeta (\kappa-\mu) + \rho)^2 + (1-\rho^2) } \right).
\end{equation}
Our SSVI parameterization of a surface corresponds to
$\mbox{SVI}_T=$ $(0, 0, \rho, \Theta_T, \phi(\Theta_T) )$
for each $T$,
where $\Theta_T$ is the at-the-money total implied variance and we use for $\phi$ a power law function $\phi(\vartheta) = \frac{\eta} {\vartheta^{\gamma} (1 + \vartheta)^{1 - \gamma}}$. \cite[Remark 4.4]{gatheral2014arbitrage} provides
sufficient conditions on $\text{SSVI}$ parameters ($ \eta (1 + |\rho|) \leq 2$ with $\gamma=0.5$) that rule out butterfly arbitrage, whereas SSVI is free of calendar arbitrage when $\Theta_T$ is nondecreasing. 

We calibrate the model as in \cite{gatheral2014arbitrage}:\footnote{Building on
%a MATLAB implementation of Philipp Rindler, 
https://www.mathworks.com/matlabcentral/profile/authors/4439546.}
First, we fit the SSVI model; Second, for each maturity in the training grid, the five SVI parameters are calibrated, (starting in each case from the SSVI calibrated values.
The \IV  is obtained for new maturities by a weighted
 average of the parameters associated with the two closest maturities in the training grid, $T$ and $U$, say, with weights determined by  $\Theta_T$ and $\Theta_U$. 
The  corresponding  local  volatility  is  extracted  by  finite  difference approximation of \ref{gatheral}.

As, in practice, no arbitrage constraints are implemented for SSVI by penalization (see  \cite[Section 5.2]{gatheral2014arbitrage}), in the end the SSVI approach is in fact only practically arbitrage-free, much like our NN approach, whereas it is only the GP approach that is proven arbitrage-free.

 \subsection{Calibration results}

Training times for SSVI, GP, and NNs are reported in the last row of Table \ref{tab:backtest} which, for completeness, also includes numerical results obtained by NN interpolation of the prices as per \cite{Chataigner2020Deep}. Because price based NN results are outperformed by IV based NN results we only focus on the IV based NN in the figures that follow, referring to \cite{Chataigner2020Deep} for every detail on the price based NN approach. We recall that, in contrast to the SSVI and NNs which fit to mid-quotes, GPs fit to the bid-ask prices.
   \begin{table}[h!]
  \centering
\resizebox{\columnwidth}{!}{
\begin{tabular}{|c|c|c|c|c|c|c|c|c|}
\hline
\begin{tabular}[c]{@{}c@{}}IV RMSE\\
%$[$*$]$ 
(Price RMSE)
%(impl.~vol RMSE) (impl.~vol RMSE reduced)
\end{tabular}             & SSVI                                                     & \begin{tabular}[c]{@{}c@{}}GP\\\end{tabular} & \begin{tabular}[c]{@{}c@{}}IV based \\  NN\end{tabular} & \begin{tabular}[c]{@{}c@{}}Price \\ based NN\end{tabular} & \begin{tabular}[c]{@{}c@{}}SSVI \\ Unconstr.\end{tabular} & \begin{tabular}[c]{@{}c@{}}GP \\ Unconstr.\end{tabular} & \begin{tabular}[c]{@{}c@{}}IV based\\  NN \\ Unconstr.\end{tabular} & \begin{tabular}[c]{@{}c@{}}Price \\ based NN \\ Unconstr.\end{tabular}  \\ \hline
\hline
\begin{tabular}[c]{@{}c@{}}Calibr. fit on \\ the training set\end{tabular}              & \begin{tabular}[c]{@{}c@{}}	1.37\% 
%\\ $[ 17.01\%]$ 
\\ (2.574) \end{tabular} &              
\begin{tabular}[c]{@{}c@{}}0.58\%
%\\ $[5.67\%]$
\\ (0.338)  \end{tabular}  &            
\begin{tabular}[c]{@{}c@{}}1.23\%
%\\ $[10.65\%]$
\\ (2.897) \end{tabular} &
\begin{tabular}[c]{@{}c@{}}13.70\%
%\\ $[616.81\%]$ 
\\ (9.851) \end{tabular} &
\begin{tabular}[c]{@{}c@{}} 1.04\%
%\\ $[8.03\%]$ 
\\ (2.691)\end{tabular} &
\begin{tabular}[c]{@{}c@{}} 0.60\%
%\\ $[5.52\%]$  
\\ (0.321)\end{tabular} &
\begin{tabular}[c]{@{}c@{}} 0.84\% 
%\\ $[12.81\%]$  
\\ (2.163) \end{tabular} &
\begin{tabular}[c]{@{}c@{}} 5.65 \%
%\\ $[128.57\%]$  
\\ (2.456) \end{tabular}\\ \hline

\begin{tabular}[c]{@{}c@{}}Calibr. fit on 
\\ the testing set\end{tabular}               & 
\begin{tabular}[c]{@{}c@{}}	1.52\%
%\\ $[17.81\%]$  
\\ (2.892) \end{tabular} & 
\begin{tabular}[c]{@{}c@{}}0.57\%
%\\ $[7.00\%]$ 
\\ (0.355)  \end{tabular}    & 
\begin{tabular}[c]{@{}c@{}} 1.29\%
%\\ $[11.27\%]$ 
\\ (2.966) \end{tabular} &
\begin{tabular}[c]{@{}c@{}}14.27\%	
%\\ $[904.50\%]$ 
\\ (10.347) \end{tabular} &
\begin{tabular}[c]{@{}c@{}} 1.09\% 
%\\ $[9.07\%]$  
\\ (2.791)  \end{tabular} & \begin{tabular}[c]{@{}c@{}}  0.57\%
%\\ $[6.81	\%]$   
\\ (0.477) \end{tabular}  & \begin{tabular}[c]{@{}c@{}}  0.86\% 
%\\ $[13.75\%]$  
\\ (2.045)  \end{tabular} &
\begin{tabular}[c]{@{}c@{}} 6.14\% 
%\\ $[153.34\%]$  
\\ (2.888)  \end{tabular}\\ \hline

\begin{tabular}[c]{@{}c@{}}MC backtest
%\\ Testing local Vol
\end{tabular}      & 
\begin{tabular}[c]{@{}c@{}}8.69\%   
%\\     $[64.26\%]$   
\\ (22.826)    \end{tabular}    & 	
\begin{tabular}[c]{@{}c@{}}19.76\%  
%\\  $[1787.78\%]$ 
\\    (74.017)     \end{tabular}  &  
\begin{tabular}[c]{@{}c@{}}2.95\%
%\\ $[25.10	\%]$ 
\\ (4.989) \end{tabular} &       
\begin{tabular}[c]{@{}c@{}}6.37\%  
%\\ $[56.92\%]$ 
\\ (11.764)     \end{tabular}  &
N/A              & 
N/A &           
N/A     & 
N/A                                   \\ \hline
\begin{tabular}[c]{@{}c@{}}CN backtest
%\\ Testing local Vol
\end{tabular}              & 
\begin{tabular}[c]{@{}c@{}} 6.88\%
%\\  $[222.96\%]$       
\\ (33.545) \end{tabular} & 
\begin{tabular}[c]{@{}c@{}}7.86\%   
%\\         $[72.96\%]$ 
\\ (35.270)  \end{tabular} & 
\begin{tabular}[c]{@{}c@{}}3.43\%
%\\$[112.28\%]$ 
\\ (11.976) \end{tabular} &  
\begin{tabular}[c]{@{}c@{}}5.56\% %\\$[61.38\%]$ 
\\ (26.785) \end{tabular} &  
N/A  & 
N/A                                                    &  
N/A  & 
N/A                                                            \\ \hline\hline

\begin{tabular}[c]{@{}c@{}} Comput. time \\ (seconds)
%\\ Testing local Vol
\end{tabular}              & 33                                                 & 856                                                                                 & 191  & 185  &  1  & 16                                                  &  76    &               229                                          \\ \hline

\end{tabular}
}
\caption{
% %\textit{$[*]$=rel.~price RMSE,
% $(*)$=implied vol 
The IV and price RMSEs of the SSVI, GP and NN approaches. Last row: computation times (in seconds).}
\label{tab:backtest}
\end{table}

The GP implementation is in Matlab whereas the SSVI and NN approaches are implemented in Python.
%That being said
On our (large) dataset, the constrained GP has the longest training time.
% due to the high number of listed options. 
Training is longer for constrained SSVI than for unconstrained SSVI because of the ensuing amendments to the optimization routine. 
There are no arbitrage violations observed  for any of the constrained methods in neither the training or the testing grid.
% although this is not guaranteed for NNs.
Unconstrained methods yield 18 violations with NN and 177 with SSVI on the testing set, out of a total of 1725 testing points, i.e. violations in 1.04\% and 10.26\% of the test nodes. The unconstrained GP approach yields constraint violations on 12.5\% of the basis function nodes $\imath h$. 
The NN penalizations $(\cale)^-$ and $(\butt)^-$ vanish identically on the penalty grid $\Omega_h$ in the constrained case, whereas in the unconstrained case their averages across grid nodes in $\Omega_h$ are
$(\cale)^-=3.91 \times 10^{-6}$ and $(\butt)^-=1.60 \times 10^{-2}$ with the IV based NN.
% , respectively $(\cale)^- =1.64 \times 10^{-2}$ and $(\butt)^-=7.59 \times 10^{-6}$ with price based NN.}

Fig.~\ref{fig:testMultiSlices}(a-b) respectively compare the fitted IV surfaces and their errors with respect to the market mid-implied volatilities, among the constrained methods. The surface is sliced at various maturities (more slices are available in the github) and the IVs corresponding to the bid-ask price quotes are also shown -- the blue and red points respectively denote training and test observations.

We generally observe good correspondence between the models and that each curve typically falls within the bid-ask spread, except for the shortest maturity contracts where there is some departure from the bid-ask spreads for observations with the lowest log-moneyness values. We see on Fig.~\ref{fig:testMultiSlices}(b) that the GP IV errors are
small and mostly less than 5 volatility points,
whereas NN and SSVI exhibit IV error that may exceed 15
volatility points. 
The green line and the red shaded envelopes respectively denote the GP MAP estimates and the posterior uncertainty bands under 100 samples per observation. 
The support of the posterior GP process assessed on the basis of 100 simulated paths
% Coincidentally, 
%  the 90\% 
% \r{uncertainty band} 
of the GP captures the majority of bid-ask quotes. The GP MAP estimate occasionally corresponds to the boundary of the support of the posterior simulation.
% \footnote{Using quantiles instead of the full  support of the simulations for GP uncertainty quantification, our MAP would actually lie outside the uncertainty bands for short maturities.}
This indicates that the posterior truncated Gaussian distribution is heavily skewed for some points, and that the MAP estimate consequently saturates the arbitrage constraints. This indicates a tension between these constraints and the calibration requirement, which cannot be fully reconciled, most likely because some of the (short maturity) data are arbitrable (they are at least illiquid and hence noisy). See notebook for location of arbitrages in the unconstrained approach.

Fig.~\ref{fig:testMultiSlices}(a-b) suggest that the
data
%observed IV surface
may exhibit arbitrage at the lowest maturities where the methods depart from the bid-ask spreads. This is further supported in Fig. \ref{fig:testMultiUnconstrainedSlices}(a-b) which shows the corresponding methods without the no-arbitrage constraints. 
In Fig. \ref{fig:testMultiUnconstrainedSlices}(a-b) we observe that the estimated IVs now fall within close proximity of the bid-ask spreads--all methods exhibit an error typically less than 5 volatility points. % less than 5\%. 
Note that the y-axis has been scaled for each plot in Fig.~\ref{fig:testMultiUnconstrainedSlices}(b) to accommodate the wide
%90\% 
uncertainty band of the posterior for the unconstrained GP. Whereas the 
%90\% 
uncertainty band of the constrained GP spanned at most 10 volatility points,
%a 10\% difference in IV, 
the 
%90\% 
uncertainty band of the unconstrained GP is an order of magnitude larger, sometimes spanning more than 100 volatility points.
%a 100\% difference in IV.

\begin{figure}[!htbp]
\centering
\begin{subfigure}{0.6\paperwidth}
    \centering
    \includegraphics[width=\linewidth,
    height=0.2\textheight]{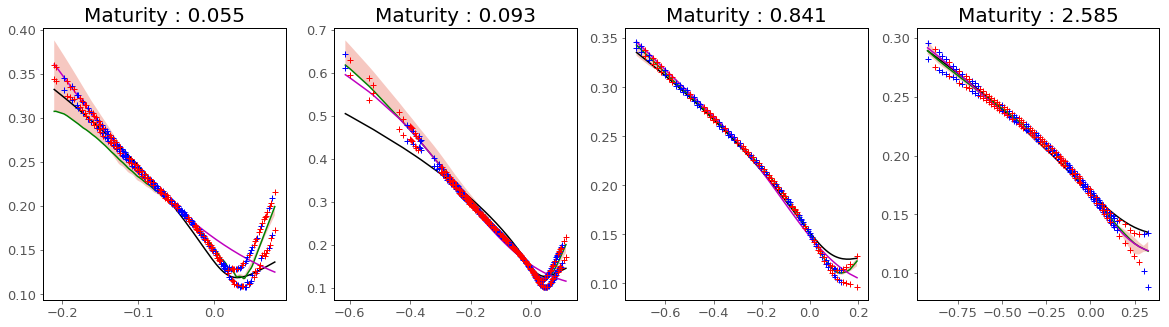}
    \subcaption{\small{Implied volatilities.}}
    \label{fig:multiIV}
\end{subfigure}
\begin{subfigure}{0.6\paperwidth}
    \centering
    \includegraphics[width=\linewidth, height=0.2\textheight]{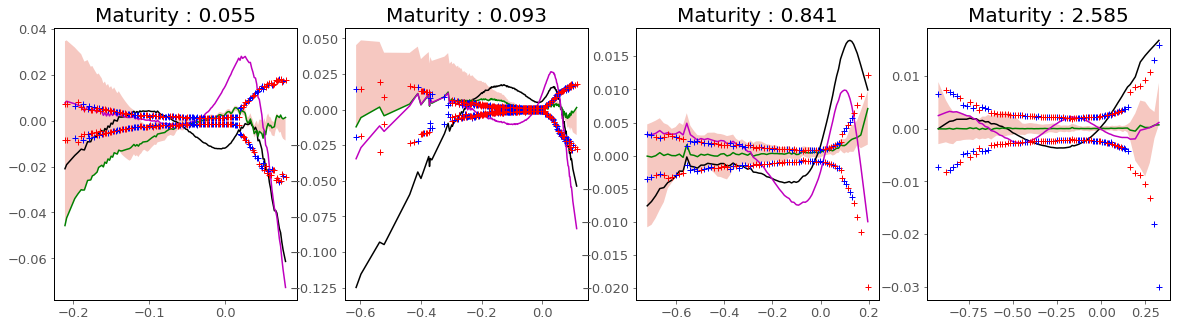}
    \subcaption{\small{Fitted IV errors with respect to mid-price IVs.}}
    \label{fig:multiPrice}
\end{subfigure}
\caption{\textit{Slices of constrained GP (green), NN (purple), and SSVI (black) models of SPX puts with training bid-asks IVs (\b{+}) and testing bid-asks IVs   as a function of log forward moneyness (\r{+})(the bid-ask IVs are reconstructed numerically from the corresponding bid-ask market prices). The shaded envelopes show 100 paths of the constrained GP's posterior.}}
\label{fig:testMultiSlices}
\end{figure}

\begin{figure}[!htbp]
\centering
\begin{subfigure}{0.6\paperwidth}
    \centering
    \includegraphics[width=\linewidth, height=0.2\textheight]{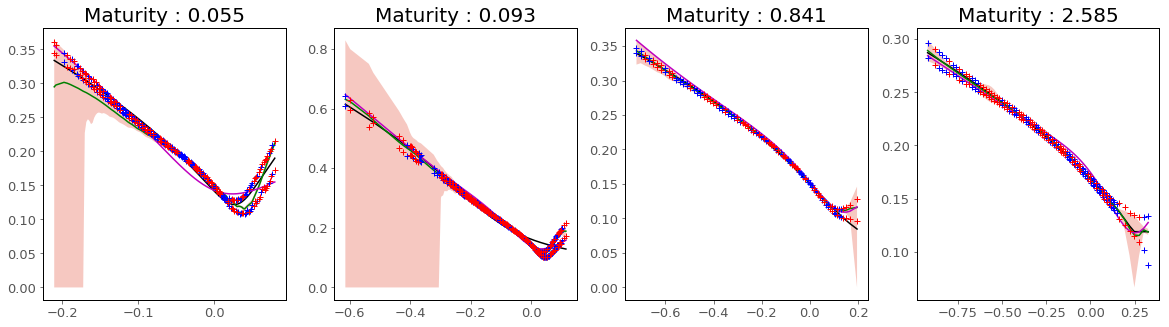}
    \subcaption{\small{Implied volatilities.}}
    \label{fig:multiUnconstrained}
\end{subfigure}
\begin{subfigure}{0.6\paperwidth}
    \centering
    \includegraphics[width=\linewidth, height=0.2\textheight]{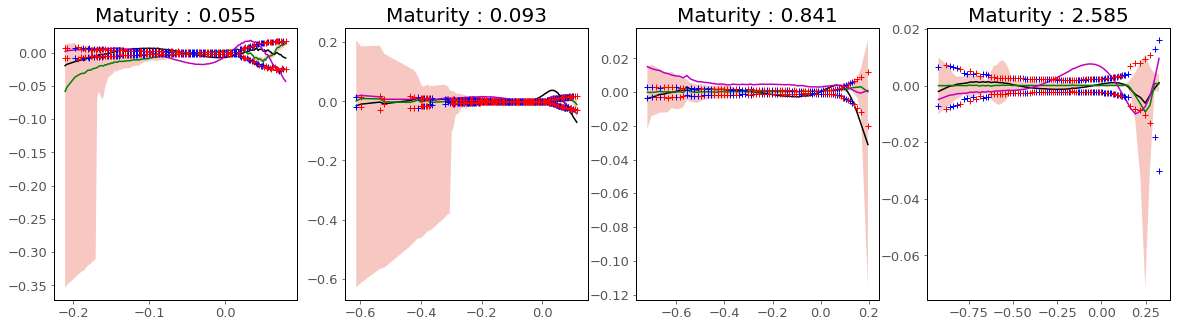}
    \subcaption{\small{Fitted IV errors with respect to mid-price IVs.}}
    \label{fig:multiUnconstrainedDiff}
\end{subfigure}
\caption{\textit{Same as Figure \ref{fig:testMultiSlices}
but for unconstrained GP, NN and SSVI.}}
\label{fig:testMultiUnconstrainedSlices}
\end{figure}
 
\begin{figure}[!htbp]
\centering
\begin{subfigure}{0.2\paperwidth}
  \centering
  \includegraphics[width=\linewidth]{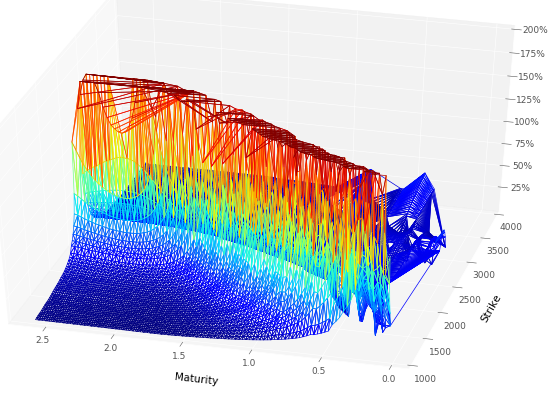} 
  \subcaption{\tiny{The local volatility surface generated by SSVI with finite differences, capped at the 200\% level.}}   
  \label{fig:SPX_GP_implied_vol_MAP}
\end{subfigure}\begin{subfigure}{0.2\paperwidth}
  \centering
  \includegraphics[width=\linewidth]{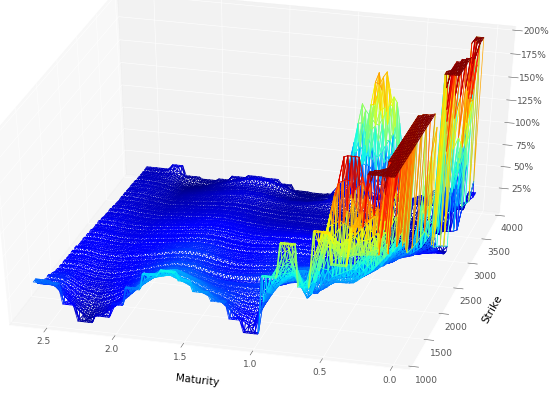}
  \subcaption{\tiny{The MAP estimate of the GP local volatility surface, capped at the 200\% level.}}
  \label{fig:SPX_GP_price_samples}
\end{subfigure}
\begin{subfigure}{0.2\paperwidth}
  \centering
  \includegraphics[width=\linewidth]{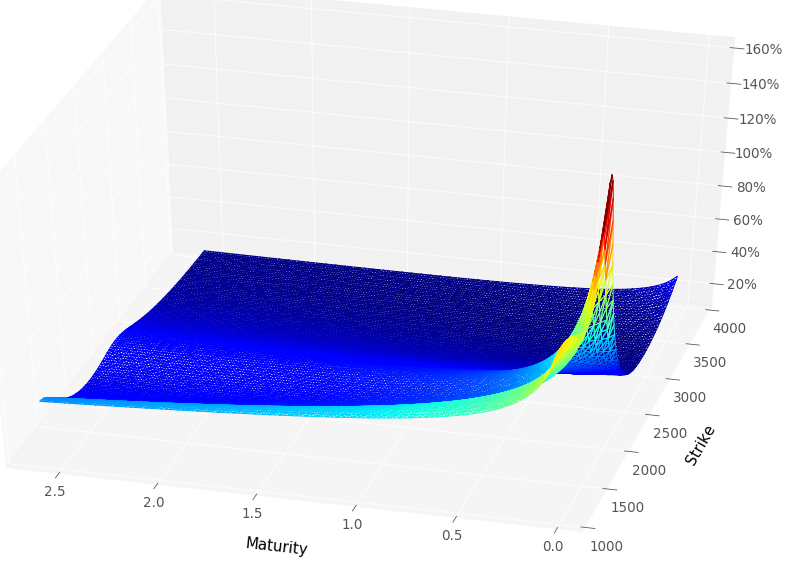}
  \subcaption{\tiny{The implied volatility based NN local volatility surface (with the local volatility penalization).}}
  \label{fig:SPX_GP_local_vol_MAP}
\end{subfigure}
\caption{\textit{The GP, SSVI, and NN local volatility estimate.}}
\label{fig:testMulti12Second}
\end{figure}

Fig.~\ref{fig:testMulti12Second} shows the  
local volatility surfaces 
that stem from the three constrained approaches.
Fig.~\ref{fig:testMulti12Second}(a) shows the spiky local volatility surface generated by SSVI, capped at the 200\% level for scaling convenience.
Fig.~\ref{fig:testMulti12Second}(b) shows the capped local volatility surface constructed from the GP MAP price estimate. Fig.~\ref{fig:testMulti12Second}(c) shows the (complete) NN local volatility surface.

\subsection{In-sample and out-of-sample calibration errors}
%In Sample vs. 
%Neural nets on implied volatilities

 The error between the prices of the calibrated models and the market data are evaluated on both the training and the out-of-sample data set. The first two rows of Table \ref{tab:backtest}
compare the in-sample and out-of-sample RMSEs of the prices and implied volatilities across the different approaches. The differences between the training and testing RMSEs are small, suggesting that all approaches are not 
%excessively 
over-fitting 
%to 
the training set. The GP exhibits the lowest price RMSEs.

\subsection{Backtesting results}
The first repricing backtest estimates the prices of the European options corresponding to the testing set, by Monte Carlo sampling in each calibrated local volatility model (same methodology as in 
 \cite[Section 7.2]{Chataigner2020Deep}). The second approach uses finite differences to price the options with the calibrated local volatility surfaces. The pricing PDEs with local volatility are discretized  using a Crank-Nicolson (CN) scheme implemented on a $100\times100$ backtesting grid. The last two rows in Table~\ref{tab:backtest} compare the resulting price backtest RMSEs across the different approaches. The NN fitted to implied volatilities exhibit significantly lower errors in the backtests, followed by NN based on prices, SSVI and GP.   
% The PDE backtest shows 
% the same ordering of the results. 
To quantify discretization error in these backtesting results (as opposed to the part of the error stemming from a  wrong local volatility), we ran the same backtests in a Black-Scholes model with 20\% volatility and the associated prices. The corresponding Monte Carlo and Crank-Nicholson
backtesting IV(price) RMSEs are
$2.90\%(1.56)$ and $0.846\%(4.10)$, confirming the significance of the above results.\\
 
 \section{Conclusion}
We approach the option quote fitting problem from two perspectives: (i) the GP approach assumes noisy data and hence the existence of a latent function. The mid-prices are not considered, rather the GP calibrates to bid-ask quotes; and (ii) the NN and SSVI approaches fit to the mid-prices under a noise-free assumption. While these two approaches are important to distinguish on theoretical grounds, in practice there are other factors which are more important for, in particular, local volatility modeling. In line with classical inverse problems theory, we find that regularization of the local volatility
%, like with our NN approach, 
is critical for backtesting performance. %and we adopt one particular form of this in our NN based approach but not in the GP approach.

\section{Acknowledgements}
he authors are thankful to Antoine Jacquier and Tahar Ferhati for useful hints regarding the SSVI method, and to an anonymous referee for stimulating comments.

 \bibliographystyle{plain}
\bibliography{main-arxiv-beyond-surrogate}
\end{document}